\documentclass[latter]{spie}
\usepackage[]{graphicx}
\usepackage{dcolumn}
\usepackage{bm}
\renewcommand{\topmargin}{0.25in}
\title{Second-Order Nonlinear Mixing of Two Modes in a Planar Photonic Crystal Microcavity}

\author{Murray W. McCutcheon\supit{a}, Jeff F. Young\supit{a},
Georg W. Rieger\supit{a}, Dan Dalacu\supit{b}, \mbox{Simon Fr\'{e}d\'{e}rick\supit{b,c},} 
Philip J. Poole\supit{b}, Geof C. Aers\supit{b}, and Robin L. Williams\supit{b,c}
\skiplinehalf
\supit{a}Dept. of Physics and Astronomy, University of B.C.,
Vancouver, Canada, V6T 1Z1; \\
\supit{b}Inst. for Microstructural Sciences, National Research Council, Ottawa, Canada, K1A OR6; \\
\supit{c}Dept. of Physics, University of Ottawa, Ottawa, Canada, K1N 6N5
}
\begin{document}
\maketitle
\begin{abstract}

Polarization-resolved second-harmonic spectra are obtained from the resonant 
modes of a two-dimensional planar photonic crystal microcavity patterned in a 
free-standing InP slab.  The photonic crystal microcavity is comprised of a single
missing-hole defect in a hexagonal photonic crystal host formed with
elliptically-shaped holes.  The cavity supports two orthogonally-polarized 
resonant modes split by 60 cm$^{-1}$.  Sum-frequency data are reported from the nonlinear
interaction of the two coherently excited modes, and the polarization 
dependence is explained in terms of the nonlinear susceptibility tensor of the 
host InP.

\end{abstract}

\section{Introduction}

Planar photonic crystals (PPCs) are ideal platforms to host engineered
optical microcavities for the study of light-matter interactions in
cavity quantum electrodynamics (QED) and quantum information research.  
Although the Q-factors of PPC microcavities have typically been much lower than
the Fabry-Perot cavities used in the pioneering cavity QED 
work~\cite{Kimble,Mabuchi}, in the last three years innovative 
momentum-space~\cite{PainterPRB,Vuck05} and double heterostructure
designs have realized Q-factors as large as 600,000.  Moreover, PPC
microcavities have three distinct advantages over their macroscopic counterparts: 
they provide mode volumes which approach the fundamental limit of a cubic 
half-wavelength in material; they allow integration with quantum dots to 
control the exciton/photon coupling; and they are based on the mature 
fabrication technology of e-beam lithography and plasma-etching, which facilitates 
higher-scale integration and coupling to waveguides.  In a quantum information context, 
PPC microcavities that incorporate artificial quantum dots can be configured
in different ways to act as single photon sources~\cite{Santori}, or ``qubits'', the basic
building blocks of quantum computers.  For example, the two states that comprise the 
qubit could be two high-Q photon modes of a microcavity that are coupled through a 
nonlinearity~\cite{Zag}, or two quantum dots that are coupled by a single cavity 
mode~\cite{Imamoglu}.

We recently reported a novel resonant scattering technique that probes coherently
excited resonant modes of PPC microcavities efficiently and independently 
of the photoluminescence (PL) from embedded quantum dots~\cite{McCutcheon}.
The data were from single-missing hole defect cavities that supported two non-degenerate
modes, each with a Q of about 500.  We also reported an example of the spectra emitted
from such modes at approximately twice the frequency
of the broadband pulse that resonantly excited both cavity modes. Four
distinct features were identified:  three centered at twice the frequencies of the
excitation pulse and the two cavity modes, and one centered at the sum of the
two distinct cavity mode frequencies.
In this paper, the polarization dependence of these harmonic features
is explored and interpreted in terms of the nonlinear
susceptibility tensor of the host InP.  The data can be explained principally
in the context of the bulk $\bar{4}3m$ $\chi^{(2)}$ 
tensor, although there may be additional surface effects arising from the
broken symmetry around the air holes that define the PPC microcavity.  Since the
sum-frequency signal can only be generated when both cavity modes are excited, this
offers a means of monitoring the occupation of the cavity modes in a weakly invasive
manner.  These results also suggest that the inverse nonlinear process, whereby a single
photon at the sum-frequency generates two photons in the microcavity modes, might
allow these to act as miniature sources of correlated photon pairs.

\section{Experiment}

The microcavity is defined in a 2D pattern of air holes in a 223 nm 
free-standing layer of InP separated by 1 $\mu$m from a glass substrate.  
A single layer of InAs quantum dots at a relatively low density of 1-2 $\mu$m$^{-2}$
is contained in the middle of the InP membrane.  The cavity-enhanced
PL from the quantum dots was reported 
previously~\cite{Dalacu05,McCutcheon}.  Further details about the sample 
fabrication can be found in Dalacu et al.~\cite{Dalacu05}

\begin{figure}[h]
\centering
 \includegraphics[width=8cm]{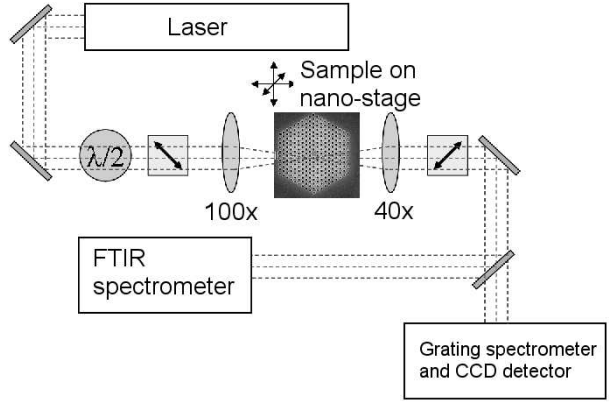}
\caption{\label{fig:setup}Experimental set-up for resonant scattering and
harmonic spectroscopy experiments.  (In reality, the sample is mounted normal 
to the plane of incidence, but is rotated in the diagram for clarity.)}
\end{figure}

The harmonic spectroscopy is performed in tandem with resonant scattering, 
as shown in the experimental set-up in Figure~\ref{fig:setup}.  
In both techniques, a Spectra-Physics optical parametric
oscillator (OPO) tuned near 1580 nm is used to generate 100 fs pulses at
a repetition rate of 80 MHz.  The desired excitation polarization is selected
using a half-wave plate and polarizer, and the light is tightly focused
by a 100x microscope objective.  It is collected in transmission using a 
40x microscope objective (NA 0.55).  With the input lens defocused, the PPC 
outline can be clearly imaged on a camera placed after the collection lens,  
and the beam focussed down to the missing-hole defect
cavity.  The resonant scattering signal, which has excellent signal-to-noise and 
is rapid to acquire with a Bomem
Fourier transform infrared (FTIR) spectrometer, is used to align the cavity for the harmonic spectroscopy.  
By optimizing this signal, the position of the sample, which is mounted on Melles 
Griot nanopositioners, can be adjusted to high precision to effectively excite each
mode separately, or both modes, depending on the incident polarization.

To acquire the second-harmonic data, average excitation powers of 70-110 $\mu$W are used, and the 
signal is directed to a grating spectrometer and cooled charge-coupled device (CCD)
detector by removing the final turning mirror.  The only adjustment to the detection 
scheme is that the collection lens must be translated about 50 $\mu$m towards the 
sample to account for the chromatic dispersion between the fundamental and 
second-harmonic frequencies.

\section{Results and Discussion}
\subsection{Off-cavity harmonic spectroscopy}

\begin{figure}[h]
\begin{center}
$\begin{array}{c@{\hspace{0.5in}}c}
\includegraphics[width=7cm]{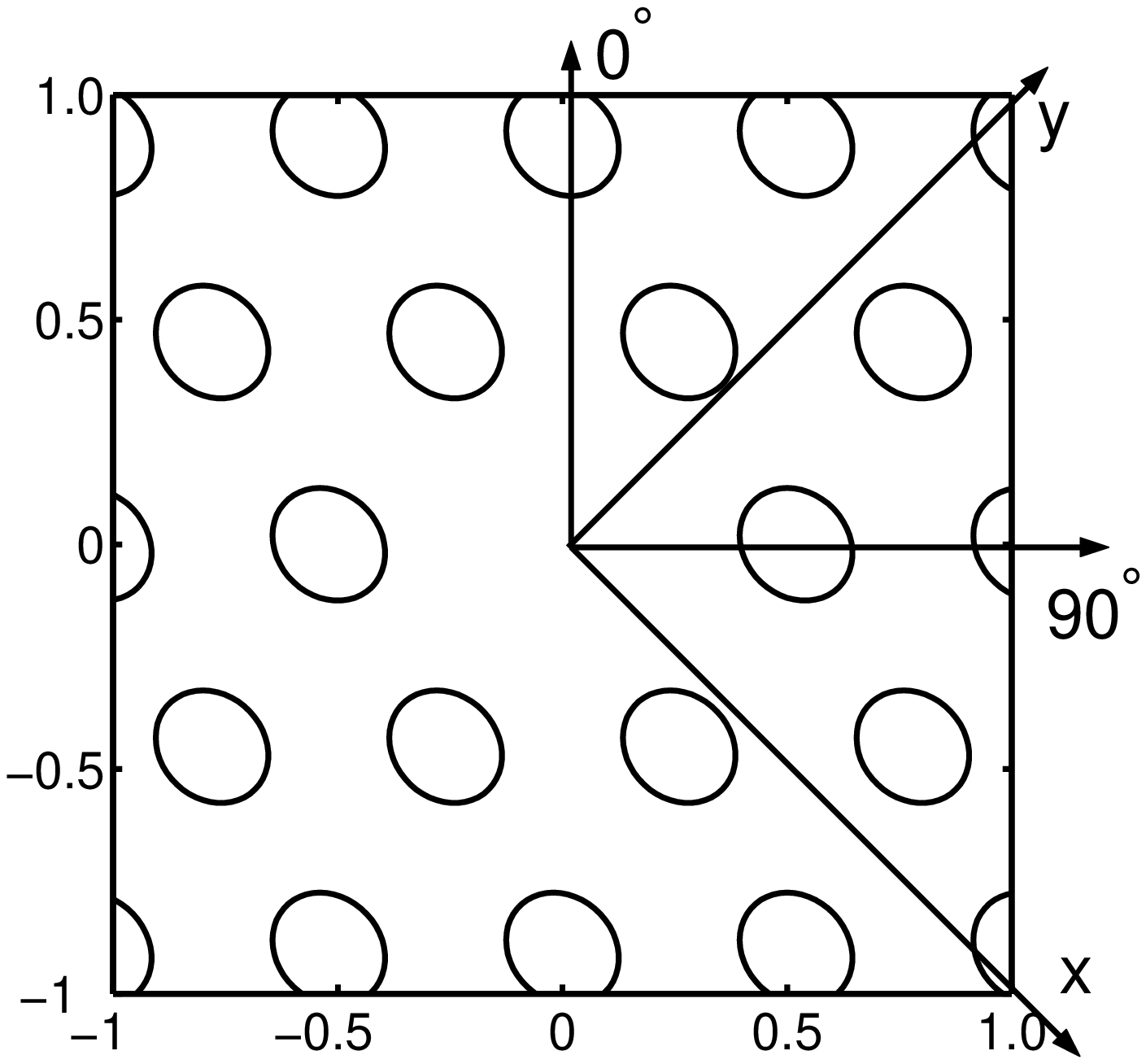}  &
\includegraphics[width=7cm]{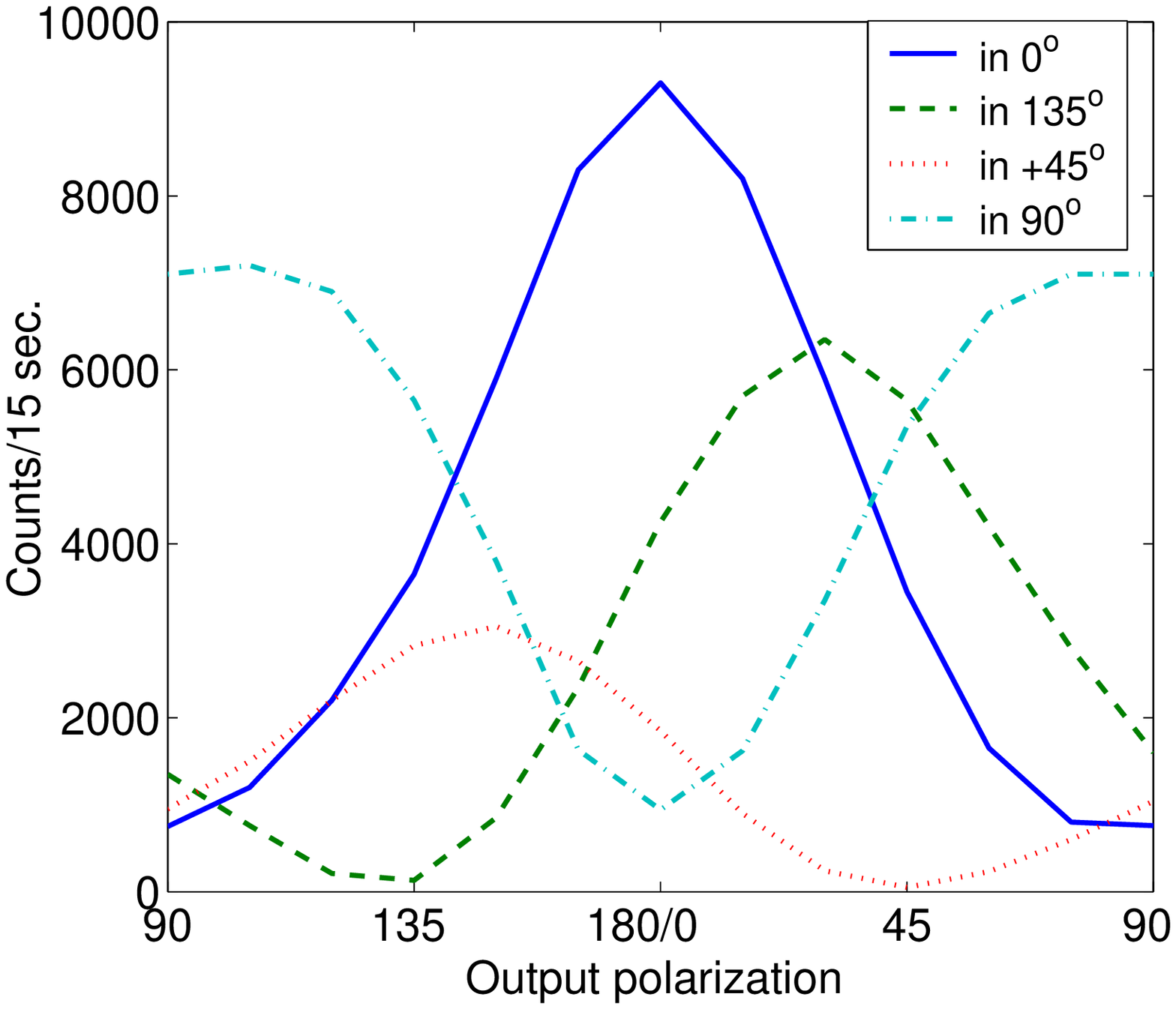} \\ 
\mbox{\bf (a)} & \hspace{0.3in}\mbox{\bf (b)}
\end{array}$
\end{center}
\caption{\label{fig:shgoff}a) Schematic of PPC microcavity geometry with 
elliptically-stretched holes, electronic crystal axes (x,y), and polarization 
reference directions
(dimensions are in microns).
b)  Second harmonic intensities for 4 different input polarizations on the unpatterned InP 
membrane.  The 45$^\circ$ and 135$^\circ$ polarizations coincide with the $y/x$ 
InP electronic axes, respectively.}
\end{figure}

To interpret the microcavity results, it is first necessary to understand the ``bulk''
membrane response.  Away from the patterned regions, the InP membrane is not free-standing,
but is supported by a 1 $\mu$m layer of SiO$_2$, which is glued with optical adhesive to a 
thick glass substrate.  The second-harmonic response of the untextured, supported
membrane was probed in each of
four incident polarizations:  0$^\circ$, 45$^\circ$, 90$^\circ$, and 135$^\circ$.  
The $y/x$ electronic axes of the InP are aligned with the 45$^\circ$/135$^\circ$ 
polarizations, respectively, as shown in Fig.~\ref{fig:shgoff}a.  
The detected intensity as a function 
of output polarization is shown in Figure~\ref{fig:shgoff}b.  Virtually identical results
were obtained from a sample of similar dimensions which did not contain a dilute
layer of InAs quantum dots.  The 45$^\circ$ and 135$^\circ$ data peak when the 
output signal is cross-polarized with respect to the input, and have a null in the 
parallel-polarization.  In contrast, the 0$^\circ$ and 90$^\circ$ data peak in the parallel 
polarization, and have a non-zero offset in cross-polarization.

The bulk nonlinear susceptibility tensor for InP, a zincblende ($\bar{4}3m$) crystal, 
has only three non-vanishing components:  $\chi^{(2)}_{ijk}, i\neq j \neq k$. 
In order to generate second-harmonic radiation from the bulk 
when the excitation beam is polarized parallel to either the $x$- or $y$-electronic axis, 
there must be a $\hat{z}$-component of the field, and such radiation would be cross-polarized
with respect to the input, as observed in the data.  The existence of longitudinal ($\hat{z}$)
fields is well known for tightly focussed Gaussian beams~\cite{Davis,Lax,Barton}, 
for which the paraxial approximation breaks down.  Near the beam waist, it can be
shown that $E_z \propto xE_x$~\cite{Davis}, so the longitudinal field is anti-symmetric
about the y-z mirror plane.  In our experiment,
we estimate the beam waist radius to wavelength ratio ($w_0/\lambda$) 
to be about 1, and finite-difference time-domain (FDTD) simulations 
(Lumerical Solutions) show that the maximum of $|E_z|$ is on the order of 6\% of the transverse 
field magnitude $|E_x|$ in this case.  Because of the odd symmetry of $E_z$, the second-harmonic
polarization which is generated will be of opposite sign on either side of the beam
centre.  Using FDTD, this situation can be modelled as two oppositely-oriented, in-plane
dipoles separated by about 1.5 $\mu$m.  The results confirm that there is a non-trivial 
far-field radiation pattern generated, and that the magnitude of the radiation within the
solid angle seen by the collection lens is consistent with the observed signal.

For 0$^\circ$ and 90$^\circ$ incident polarizations, both $\hat{x}$- and $\hat{y}$-polarizations
are excited in the InP slab.  The large signal in the parallel orientation and non-zero
signal in the crossed orientation are consistent with the bulk $\chi^{(2)}$
response in which the three non-vanishing tensor elements are equal in magnitude.  
The second-order polarization generated with these incident polarizations can be 
expressed as
$\vec{P} \propto \chi^{(2)}_{xyz}E_y E_z\hat{x} + \chi^{(2)}_{yxz}E_x E_z\hat{y} + 
\chi^{(2)}_{zxy}E_x E_y\hat{z}$.   
When the incident field is oriented at $\theta=0^\circ, E_x=E_y$, and the in-plane component of 
the polarization will then be oriented parallel or anti-parallel to the incident field, 
depending on the orientation of $E_z$.  The resulting second-order field radiated by the 
in-plane components of the polarization will thus be oriented parallel to the incident field.  
There will also be a $\hat{z}$-oriented dipole that radiates with no preferred polarization, which 
explains the non-zero background in this polarization configuration.

In such thin membranes, it is important to consider the role of surface second-harmonic
generation (SHG).  This is a commonly used probe of elemental semiconductors~\cite{Shen}, 
for which the bulk $\chi^{(2)}$ tensor is zero due to centrosymmetry, 
but has also been applied to non-centrosymmetric crystals~\cite{Stehlin,Yamada,Hollering}.  
For a zincblende surface, which has symmetry $4$mm, the non-zero 
surface symmetry susceptibility elements are $\chi_{zzz}, \chi_{zii}$, 
and $\chi_{izi}$, where $i=x,y$.  The latter element would give rise to radiation 
polarized parallel to the excitation field.  However, the 45$^\circ$ and 135$^\circ$
data show a nearly complete null in the parallel polarization, indicating little
sensitivity to surface effects in the normal incidence geometry of this experiment.

\subsection{Microcavity harmonic spectroscopy}

\begin{figure}[h]
\centering
 \includegraphics[width=9cm]{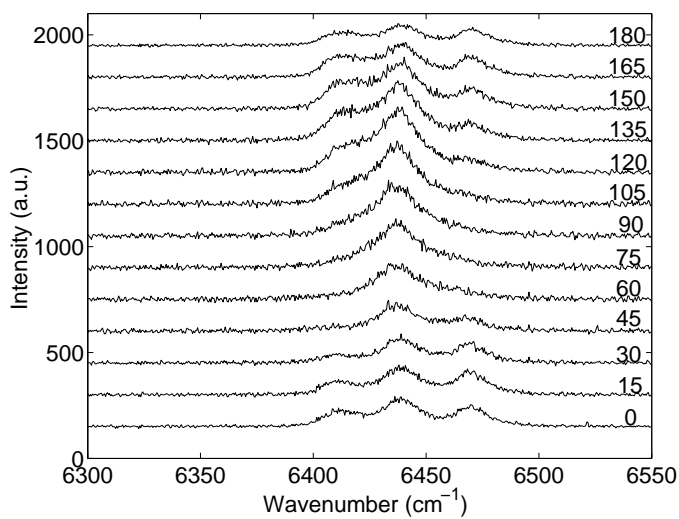}
\caption{\label{fig:shg90}Harmonic spectra with the laser excitation 
polarized at 90$^\circ$, which excites both cavity modes.  The output polarization
is indicated at the right of each trace.  There are SHG features
from both modes and the laser, and a sum-frequency peak from the nonlinear mixing of 
the two modes.}
\end{figure}

With the laser focused on the microcavity by optimizing on the modes observed
in resonant scattering, second-harmonic data were acquired at the same four 
input polarizations as for the unpatterned region.  At the 45$^\circ$ and 135$^\circ$ 
polarizations, where the laser is approximately aligned with one of the cavity modes, 
there are two features:  a broad 
peak which corresponds to the second-harmonic of the laser spectrum, and a sharp 
feature at the second-harmonic frequency of the lower energy resonant mode.  
The picture is more complicated when the laser is polarized at $0^\circ$ or
$90^\circ$.  The spectra exhibit features due to the second-harmonic of each mode, the 
sum-frequency generation from the mixing of the two modes, and the broad 
second-harmonic of the laser spectrum.  The $90^\circ$ polarization-resolved 
data are shown in Figure~\ref{fig:shg90} (the $0^\circ$ spectra are similar).

Unlike the PL spectroscopy, which imparts no coherent relation between the two modes; 
resonant, short-pulse excitation enforces well-defined amplitude and phase relations 
between the modes and the laser fields.  In the linear resonant scattering 
experiment~\cite{McCutcheon}, this was shown to lead to a
$\pi$ phase shift between the modes, as detected in the crossed polarization.  
In the present context, the data are therefore fit using the square of a coherent sum
of three Lorentzians and a laser lineshape function. 

The observation of strong SHG signals from the individual modes, polarized parallel to
the mode polarization, cannot be explained using a bulk second-order polarizability.  Because the
modes are TE-like, they are symmetric about the mirror plane in the middle of the waveguide, 
and so there should be considerable cancellation (from the top and bottom halves of the slab)
due to polarizations derived from the 
tensor elements $\chi^{(2)}_{xyz}$ or $\chi^{(2)}_{yxz}$.  Also, because the individual modes 
are approximately aligned along the electronic axes of the InP, their self-interaction 
through the $\chi^{(2)}_{zxy}$ tensor element should be negligible, and even if such a second-order
polarization were generated, it would not be polarized parallel to the mode polarization
(as in the data).  The SHG signals observed from the individual modes are therefore 
attributed to a second-order polarization generated in the vicinity of the etched holes,
where the bulk symmetry is likely not applicable.

\begin{figure}[h]
\begin{center}
$\begin{array}{c@{\hspace{0.6in}}c}
\includegraphics[width=7cm]{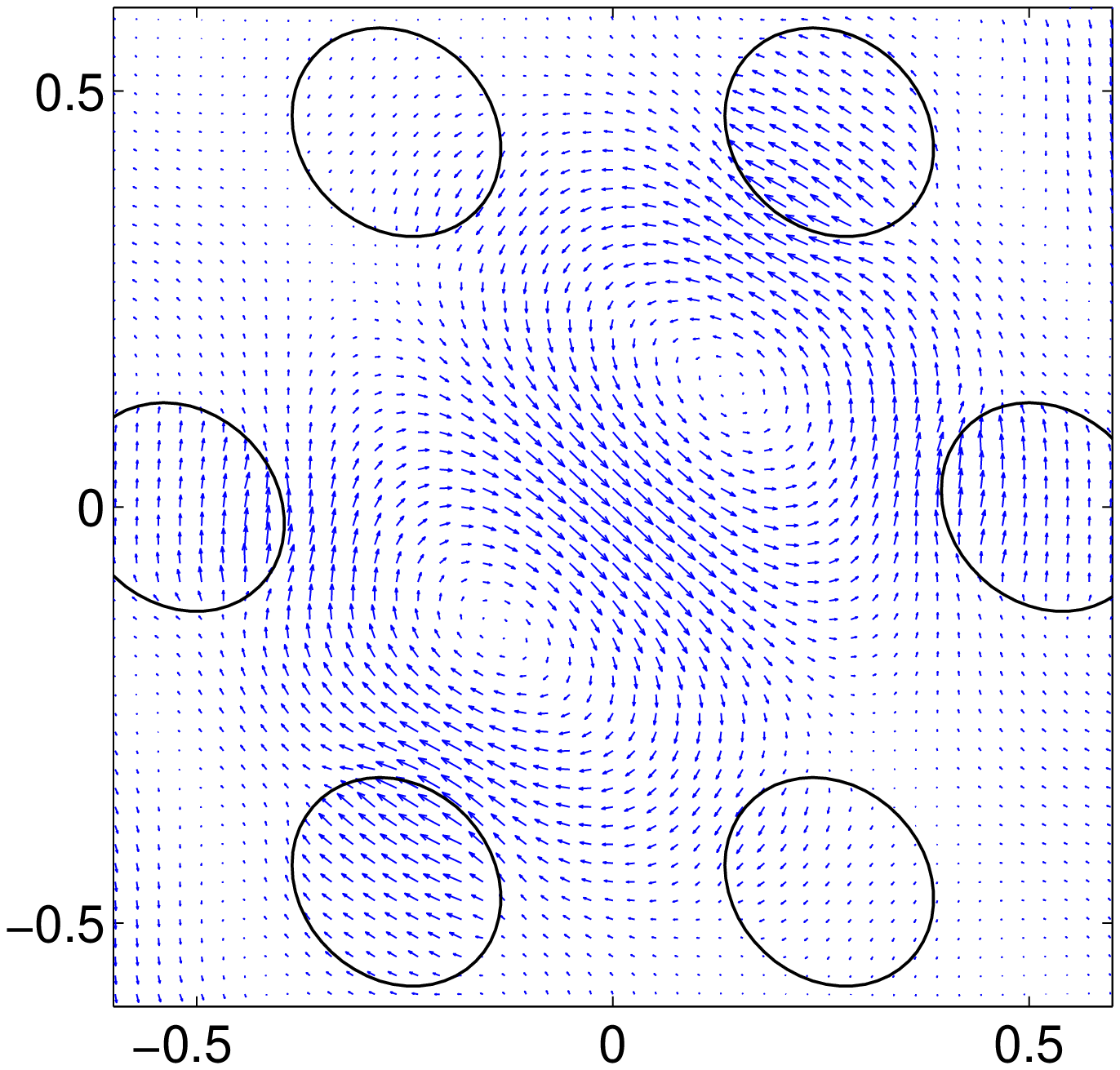}  &
\includegraphics[width=7cm]{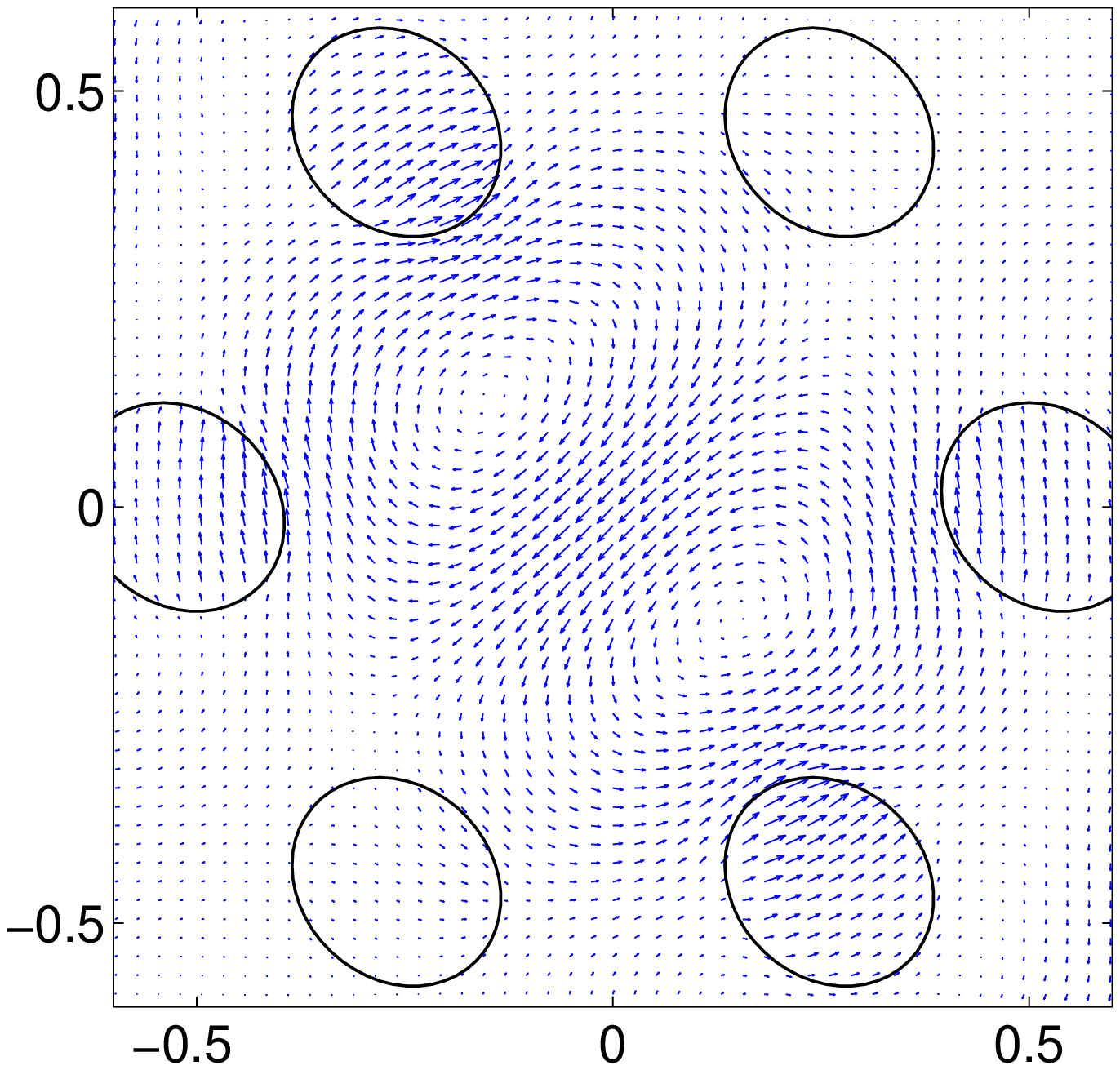} \\ 
\hspace{0.2in} \mbox{\bf (a)} & \hspace{0.2in} \mbox{\bf (b)}
\end{array}$
\end{center}
\caption{\label{fig:vec}Vector field plots of the two PPC cavity modes from FDTD
simulations, shown superimposed on the elliptically-stretched air holes defining the 
cavity.  The polarizations are approx. 135$^\circ$ (a) and 45$^\circ$ (b), consistent
with the experiment.}
\end{figure}

The experiment has been modelled by an FDTD simulation for a cavity with elliptical holes
oriented at an angle of 45$^\circ$, for which the semi-major axis is 40 nm longer than 
the semi-minor axis, which is consistent with scanning electron microscope images of the
sample.  The results give two dipole-like microcavity modes split by about 38 cm$^{-1}$, and 
polarized at approximately 135$^\circ$ (mode 1) and 45$^\circ$ (mode 2).  Although
the experimental splitting is larger, the simulation qualitatively captures the observed
mode orientation.  Vector field plots for the two modes superimposed on a schematic of
the cavity geometry are shown in Figure~\ref{fig:vec}.  From these plots it is clear that 
considerable field intensity is localized near the air holes, and so it is possible that 
new tensor components due to surface effects may be responsible for the generation
of the SHG signal from the individual modes, and also for their anomalous polarization 
dependencies.

\begin{figure}[h]
\centering
 \includegraphics[width=8cm]{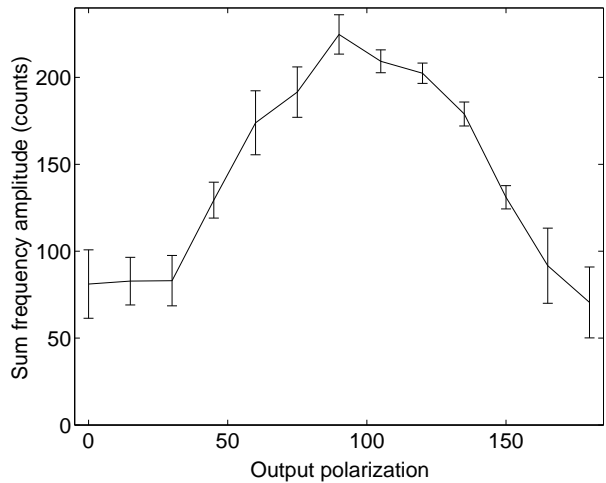}
\caption{\label{fig:pzfits}Sum-frequency amplitude for the 90$^\circ$ input
polarization data.  The error bars are derived from the standard deviations
extracted from the fit results.  The non-zero offset arises from a 
$\hat{z}$-polarization generated from the nonlinear mixing of the two
microcavity modes with a bulk $\chi^{(2)}$ symmetry.}
\end{figure}
The polarization dependence of the sum-frequency signal is shown in 
Fig.~\ref{fig:pzfits}, where the amplitude and standard deviations have been extracted 
from the nonlinear least-squares fit.  We believe that 
the non-zero offset indicates a significant component 
of the radiation derives from a nonlinear electronic polarization in the InP which lies 
along the optic ($\hat{z}$) axis.  This ``intrinsic'' response
is consistent with the orthogonally-oriented modes 
interacting through the $\chi^{(2)}_{zxy}$ component of the bulk nonlinear susceptibility 
to produce a $\hat{z}$-polarization in the sample.  The modulation in the signal, which is
also significant, is likely an ``extrinsic'' effect associated with fields localized 
near the surfaces of the etched holes, as discussed earlier.

\begin{figure}[h]
\centering
 \includegraphics[width=9cm]{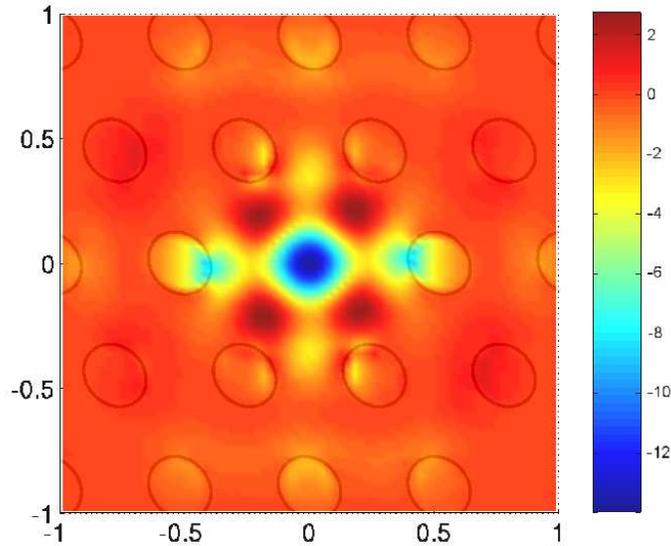}
\caption{\label{fig:pzs}FDTD simulation of the sum-frequency polarization
$P(\omega_1 + \omega_2,\vec{r})= \epsilon_0 \chi^{(2)}_{zxy} 
(E_x(\omega_1,\vec{r})E_y(\omega_2,\vec{r})+E_y(\omega_1,\vec{r}) E_x(\omega_2,\vec{r}))$ 
in the middle of the slab.  
The scalebar (arbitrary units) shows that the polarization is concentrated at the center 
of the defect cavity, where the bulk $\chi^{(2)}$ should dominate.}
\end{figure}

To explain why the bulk $\chi^{(2)}$ properties seem to play an important role in the 
sum-frequency mixing of the two modes, the spatial dependence of the sum-frequency polarization, 
$P_z(\omega_1+\omega_2,\vec{r})$, can be reconstructed from the FDTD simulation, 
as shown in Fig.~\ref{fig:pzs}.  In the limit that the modes are spectrally well-separated
and can be represented by harmonic amplitudes, the sum-frequency polarization is given
by $P_z(\omega_1+\omega_2,\vec{r})\,=\,
\epsilon_0 \chi^{(2)}_{zxy} (E_x(\omega_1,\vec{r}) E_y(\omega_2,\vec{r}) + 
E_y(\omega_1,\vec{r}) E_x(\omega_2,\vec{r})) $. 
The image in Fig.~\ref{fig:pzs} shows that the 
sum-frequency polarization derived in this manner is localized predominantly at the 
center of the cavity, where the bulk nonlinear response would be expected to dominate,
based on the off-pattern results discussed above.

Finally, a quantitative check on this picture is provided by simulating the radiated power from 
a $\hat{z}$-oriented dipole, scaled to match the strength of the polarization in the cavity, 
which is collected by a large NA lens oriented in the 
$\hat{z}$-direction, as in the experiment.  The results of this simulation are
consistent with the measured sum-frequency signal.   

\section{Conclusions}
This work demonstrates that second-order nonlinear mixing of coherently-excited
resonant modes in an InP-based PPC microcavity is detectable, and that the bulk 
nonlinear polarizability of the InP is responsible for a significant component of the 
measured signal.  Further work is needed to determine the origin of the SHG signal from 
the individual cavity modes, and the source of the polarization modulation in the 
case of the sum-frequency signal, both of which are likely associated with surface
effects at the boundaries of the etched holes used to define the microcavities.

These results motivate further studies of nonlinear interactions in multi-mode
PPC microcavities.  In particular, higher Q modes with frequencies less than
half the bandgap of the host material may make PPC-based entangled photon sources
possible.  More generally, it will be interesting to quantify the ways in which 
resonant nonlinearities, associated with excitonic transitions in embedded quantum dots, 
can influence the distribution of the field between different high-Q, small volume 
microcavity modes.




\subsection*{Acknowledgements}

The authors  wish to  acknowledge the financial support of the Natural Sciences and
Engineering Research Council of Canada, the Canadian Institute for Advanced Research,
the Canadian Foundation for Innovation, and the technical assistance of Lumerical 
Solutions Inc.


\bibliography{shgpaper}

\end{document}